\documentclass[preprintnumbers,amsmath,amssymb,prb,
singlecolumn]{revtex4}

\usepackage{graphicx}   
\usepackage{dcolumn}    
\usepackage{bm}         

\begin{document}

\title{Spectroscopy of double quantum dot two-spin states by tuning the
inter-dot barrier}

\author{G. Giavaras}
\affiliation{Faculty of Pure and Applied Sciences, University of
Tsukuba, Tsukuba 305-8571, Japan}

\author{Y. Tokura}
\affiliation{Faculty of Pure and Applied Sciences, University of
Tsukuba, Tsukuba 305-8571, Japan}



\begin{abstract}
Transport spectroscopy of two-spin states in a double quantum dot
can be performed by an AC electric field which tunes the energy
detuning. However, a problem arises when the transition rate
between the states is small and, consequently, the AC-induced
current is suppressed. Here, we show that if the AC field tunes
the inter-dot tunnel barrier then for large detuning the
transition rate increases drastically resulting in high current.
Multi-photon resonances are enhanced by orders of magnitude. Our
study demonstrates an efficient way for fast two-spin transitions.
\end{abstract}

\maketitle

\section{Introduction}

A double quantum dot containing two electron spins can be used for
the realization of two-qubit operations.~\cite{hanson07,
zwanenburg13, petta05, loss98, brunner11, pioro08} This qubit
system is attractive because it is tunable by applying appropriate
voltages to surface gate electrodes.~\cite{hanson07, zwanenburg13}
Information on the coupled spin-qubits can be extracted by
electrical transport techniques.~\cite{hanson07, zwanenburg13} It
has been demonstrated that in double dots with spin-orbit
interaction (SOI) transitions between the two-spin eigenstates can
be induced by applying an AC electric field to a gate
electrode.~\cite{perge12, stehlik14, ono17} In essence the AC
field changes periodically the energy detuning of the double dot,
and the transitions give rise to current peaks when the condition
$n h f\approx \Delta E$ is satisfied. The energy splitting of the
relevant double dot eigenstates is $\Delta E$, the AC frequency is
$f$, Planck's constant is $h$, and the integer $n$ denotes the
`$n$-photon' resonance. This process enables spectroscopy of the
spin states by measuring the current as a function of the AC field
frequency and magnetic field.~\cite{perge12, stehlik14, ono17}

The amplitude of the AC electric field is an important parameter,
because if it is small the transitions are slow and, consequently
the AC-induced current peaks are suppressed and eventually cannot
be probed. This is more evident for the peaks corresponding to
$n$-photon resonances ($n>1$) which are usually visible only when
the AC field amplitude is large.~\cite{stehlik14, ono17} However,
generating a large AC amplitude is a challenging
task.~\cite{wiel02} This is one of the basic reasons that
$n$-photon resonances are not well studied in semiconductor
quantum dot systems.

In this work, we consider a double dot (DD) in the regime of spin
selective inter-dot tunneling,~\cite{ono00} and in the presence of
an AC electric field. We focus on two SOI-coupled eigenstates, and
show that when the AC field tunes the inter-dot tunnel coupling
(barrier) the singlet-triplet~\cite{com1} transitions can be over
an order of magnitude faster compared to those induced when the AC
field tunes the energy detuning. This speedup is attractive for
spin-qubit applications where fast operations are needed. The long
time scale difference has a strong impact on the AC-induced
current peaks. Specifically, the $n$-photon resonances which are
usually suppressed and are more difficult to probe are enhanced by
orders of magnitude. As a result, the required AC field amplitude
can be smaller.

Understanding the effect of tuning periodically the tunnel
coupling of a DD is important from a more general perspective. In
impurity-based DDs such that formed, for example, in silicon
field-effect-transistors,~\cite{ono17, ono18} the position of the
dots inside the channel is unknown and the effect of applying an
AC field to a gate electrode on the dot potential is unclear. The
AC field can modulate the energy detuning and/or the inter-dot
tunnel coupling. This scenario has also been suggested in gated
quantum dots~\cite{hanson07, zwanenburg13, nakajima18}, while
control over the inter-dot tunnel coupling has been explored in
various experiments.~\cite{diepen18,mukhopa18,martins16,reed16}

\section{Double dot Hamiltonian}

We define the triplet states $|T_{\pm,0}\rangle$ as well as the
singlet states $|S_{nm}\rangle$ where $n$ ($m$) indicates the
number of electrons on dot 1 (dot 2). The DD is described by the
Hamiltonian
\begin{equation}\notag
\begin{split}
&H_{\mathrm{DD}}=\Delta[|T_-\rangle\langle T_-|-|T_+\rangle\langle
T_+|]-\delta|S_{02}\rangle\langle
S_{02}|\\
&-\sqrt{2}T_{\mathrm{c}}|S_{11}\rangle\langle S_{02}| -
T_{\mathrm{so}}[|T_+\rangle\langle S_{02}|+ |T_-\rangle\langle
S_{02}|]+ \text{H.c.}
\end{split}
\end{equation}
The Zeeman term is $\Delta=g \mu_{B} B$, with $g=2$, the tunnel
coupling is $T_{\mathrm{c}}$, the spin-flip tunnel coupling due to
the SOI is $T_{\mathrm{so}}$, and the detuning is $\delta$. We
consider two SOI-coupled eigenstates and demonstrate that by
tuning the inter-dot tunnel coupling with an AC field creates
strong $n$-photon current peaks which allow for spectroscopy of
these eigenstates at smaller AC frequencies. We prove the efficacy
of this process through a comparison with the case where the AC
field tunes the energy detuning. Below we address two separate
cases. In the first case, the detuning is time periodic due to the
AC field
\begin{equation}\notag
\delta(t) = \varepsilon+A\sin(2\pi f t).
\end{equation}
The second term accounts for the AC field which has amplitude $A$
and frequency $f$. In the first case, the tunnel-couplings are
constant in time $T_{\mathrm{c}} = t_{\mathrm{c}}$ and
$T_{\mathrm{so}} = t_{\mathrm{so}}$. In the second case, the AC
field tunes the tunnel barrier, thus, the tunnel couplings are
\begin{equation}\notag
\begin{split}
T_{\mathrm{c}}(t) &= t_{\mathrm{c}}+A\sin(2\pi f t),\\
T_{\mathrm{so}}(t)&=t_{\mathrm{so}}+x_{\mathrm{so}}A\sin(2\pi f
t).
\end{split}
\end{equation}
In the second case, the energy detuning is constant in time
$\delta = \varepsilon$. Some experiments~\cite{ono17, stehlik14,
chorley11} have demonstrated that $t_{\mathrm{so}}<t_{\mathrm{c}}$
and we assume that the AC field satisfies this condition and take
$x_{\mathrm{so}}=t_{\mathrm{so}}/t_{\mathrm{c}}$, so the ratio
$T_{\mathrm{so}}(t) / T_{\mathrm{c}}(t)$ is constant in time. To
compare theoretically the two cases we assume the same amplitude
$A$ and only briefly consider different amplitudes. In the
experiment the required gate voltage to generate $A$ can be
different for the two cases and sensitive to many factors, such
as, the geometry of the DD, the electrode design, as well as the
material, and screening effects.

\begin{figure}
\includegraphics[width=8.0cm, angle=270]{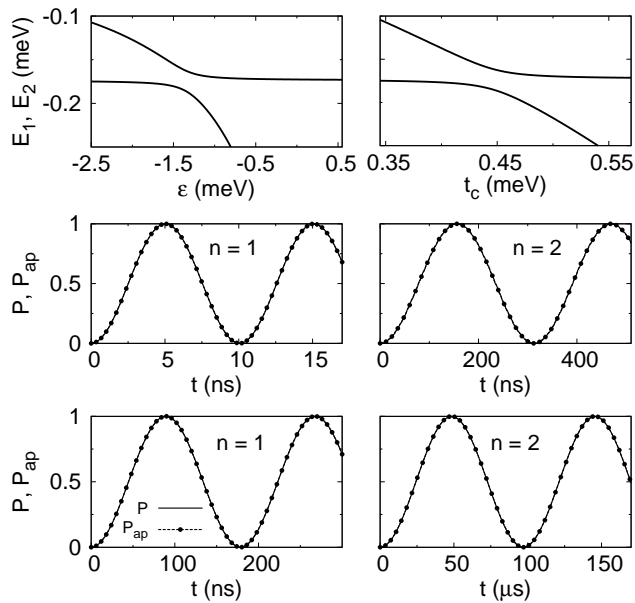}\\
\caption{The upper frames show the singlet, triplet levels
$E_{1}$, $E_{2}$ as a function of detuning (left), and tunnel
coupling (right). The middle (lower) frames show the
singlet-triplet transition probability as a function of time when
the AC field tunes the tunnel coupling (detuning). Results for the
$n=1$, 2 $n$-photon resonance are shown for $A=4$ $\mu$eV,
$\varepsilon=-2.5$ meV, $t_{\mathrm{c}}=0.35$ meV, and
$t_{\mathrm{so}}=0.035$ meV. $P$ is the exact Floquet result, and
$P_{{\mathrm{ap}}}$ is the approximate result.}\label{trans}
\end{figure}

\section{Singlet-triplet transitions}

We focus on the two lowest eigenstates of the time independent
part of $H_{\mathrm{DD}}$, $|\psi_{i}\rangle =
\alpha_{i}|S_{11}\rangle + \beta_{i}|T_{+}\rangle + \gamma_{i}
|S_{02}\rangle + \zeta_{i} |T_{-}\rangle$, $i=1$, 2, and refer to
$|\psi_{i}\rangle$ as singlet and triplet eigenstates. The
eigenenergies $E_{i}$ shown in Fig.~1 (upper frames) anticross due
to the SOI; specifically we take $B=1.5$ T,
$t_{\mathrm{so}}=0.1t_{\mathrm{c}}$ and plot $E_{1}$, $E_{2}$
versus $\varepsilon$ for $t_{\mathrm{c}}=0.35$ meV, and also plot
$E_{1}$, $E_{2}$ versus $t_{\mathrm{c}}$ for $\varepsilon=-2.5$
meV [Ref.~\onlinecite{extranote}]. Hereafter, we fix
$t_{\mathrm{c}}=0.35$ meV, $t_{\mathrm{so}}=0.035$ meV and study
first the coherent transitions between $|\psi_{1}\rangle$ and
$|\psi_{2}\rangle$, when the DD is not coupled to the leads. We
express the time evolution of the DD state as follows
\begin{equation}\notag
|\Psi(t)\rangle = \sum^{5}_{i=1} s_{i} \exp(-i \kappa_i t/\hbar )
|u_{i}(t)\rangle,
\end{equation}
where the parameters $s_{i}$ are determined by the initial
condition. The Floquet modes $|u_{i}(t)\rangle$, and Floquet
energies $\kappa_i$ satisfy the Floquet eigenvalue problem $[
H_{\mathrm{DD}}(t) - i\hbar \partial_{t} ] |u_i(t)\rangle =
\kappa_i |u_{i}(t)\rangle$. This is written in a simple form using
Fourier expansion by noting that the Floquet modes have the same
periodicity as that of the AC field, i.e.,
$|u(t)\rangle=|u(t+T)\rangle$, $T=1/f$, and the Floquet energies
can be defined within the energy interval $[-hf/2, hf/2)$. The
Floquet problem is reduced to a matrix eigenvalue problem and is
solved numerically.

Figure~1 shows the transition probability $P(t)=|\langle \psi_{2}
|\Psi(t)\rangle|^2$, $|\Psi(0)\rangle=|\psi_{1}\rangle$ for
$B=1.5$ T, $A=4$ $\mu$eV, $\varepsilon=-2.5$ meV, and $n h f
\approx\Delta E = E_{2}-E_{1}$, $n=1$, 2. This set of parameters
does not lead to Landau-Zener dynamics, as that studied in
Ref.~\onlinecite{stehlik16}, because the system is not driven
through the anticrossing point. This is inferred directly from the
dependence of $E_{1}$, $E_{2}$ on the detuning and tunnel coupling
(Fig.~1). When the AC field tunes the tunnel coupling the
singlet-triplet transition for the single-photon resonance ($n=1$)
is about 18 times faster compared to the case where the AC field
tunes the detuning. Interestingly, for the two-photon resonance
($n=2$) the speedup is on the order of 300. In both cases the
transitions are electrically driven mediated by the SOI.

In the weak driving regime transitions take place only between
$|\psi_{1}\rangle$, $|\psi_{2}\rangle$ and when the AC field tunes
the tunnel coupling the dynamics is captured by the approximate
Hamiltonian~\cite{suppl} $W = -(\Delta E-n h f)/2 \sigma_z +
q_{\mathrm{b}} \sigma_x$. This model predicts the resonant
transition probability $P_{{\mathrm{ap}}}(t) = \sin^{2}(2\pi
q_{{\mathrm{b}}} t / h)$, with
\begin{equation}\label{coupling}
q_{\mathrm{b}} = \frac{n h f
h^{\mathrm{b}}_{12}}{(h^{\mathrm{b}}_{11} - h^{\mathrm{b}}_{22})}
J_{n}\left(\frac{A(h^{\mathrm{b}}_{11} - h^{\mathrm{b}}_{22})}{hf}
\right),
\end{equation}
\begin{equation}\label{hijb}
\begin{split}
h^{\mathrm{b}}_{ij} =& - \gamma_j ( \sqrt{2}\alpha_i +
x_{\mathrm{so}} \beta_i + x_{\mathrm{so}} \zeta_i )\\ &- \gamma_i
( \sqrt{2}\alpha_j + x_{\mathrm{so}} \beta_j + x_{\mathrm{so}}
\zeta_j ), \quad i, j=1, 2
\end{split}
\end{equation}
and $J_{n}$ is a Bessel function of the first kind, $n=1$, 2, ...
denotes the `photon' index. Likewise the case where the AC field
tunes the energy detuning can be examined with the substitutions
$q_{\mathrm{b}} \rightarrow q_{\mathrm{d}}$ and
$h^{\mathrm{b}}_{ij} \rightarrow h^{\mathrm{d}}_{ij}$, where
\begin{equation}\label{hijd}
\begin{split}
h^{\mathrm{d}}_{ij} =-\gamma_i\gamma_j, \quad i, j=1, 2
\end{split}
\end{equation}

According to the approximate model the singlet-triplet transition
times are quantified by the coupling constants $q_{\mathrm{b}}$,
$q_{\mathrm{d}}$, and these can be very different. For example, if
we focus on $n=1$ and when the argument of $J_{1}$ is very small
[$J_{1}(x) \approx x/2$, $x\ll 1$] then to a good approximation
$q_{\mathrm{b}} / q_{\mathrm{d}} \approx h^{\mathrm{b}}_{12} /
h^{\mathrm{d}}_{12}$ [Ref.~\onlinecite{absol}]. For large negative
detuning where the two spins are effectively in the Heisenberg
regime $h^{\mathrm{d}}_{12} \ll h^{\mathrm{b}}_{12}$, because the
$|S_{11}\rangle$ character dominates significantly over the
$|S_{02}\rangle$, hence, near the anticrossing point $\gamma_1
\gamma_2 \ll \gamma_1 \alpha_2+\gamma_2 \alpha_1$, while the terms
proportional to $x_{\mathrm{so}}$ have negligible effect. Away
from the anticrossing point the terms proportional to $\beta_{1}$,
$\beta_{2}$ need to be included. The ratio
$q_{\mathrm{b}}/q_{\mathrm{d}}$ can be larger for $n>1$ compared
to $n=1$, thus a greater speedup can be achieved for the
$n$-photon resonances. Specifically, for the $n$-photon resonances
$q_{\mathrm{b}} / q_{\mathrm{d}} \approx [ ( h^{\mathrm{b}}_{11} -
h^{\mathrm{b}}_{22})
     / (h^{\mathrm{d}}_{11} -
h^{\mathrm{d}}_{22}) ]^{n-1} h^{\mathrm{b}}_{12} /
h^{\mathrm{d}}_{12}$, with $|h^{\mathrm{b}}_{11} -
h^{\mathrm{b}}_{22}| > |h^{\mathrm{d}}_{11} -
h^{\mathrm{d}}_{22}|$.

\begin{figure}
\includegraphics[width=9.5cm, angle=270]{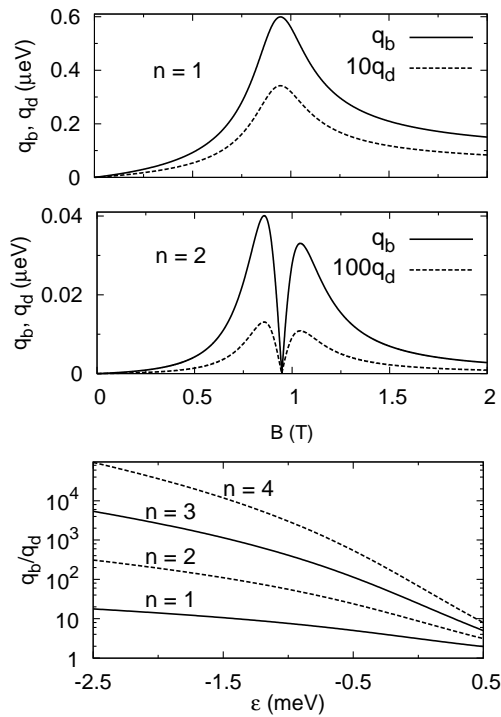}\\
\caption{The upper and middle frames show the coupling constants
$q_{\mathrm{b}}$ and $q_{\mathrm{d}}$ as a function of magnetic
field, for $n=1$, 2 $n$-photon resonance. The lower frame shows
$q_{\mathrm{b}}/q_{\mathrm{d}}$ on logarithmic scale as a function
of detuning.}\label{qdqb}
\end{figure}

The two upper frames of Fig.~\ref{qdqb} show the couplings
$q_{\mathrm{b}}$, $q_{\mathrm{d}}$ [Ref.~\onlinecite{absol}] as a
function of the magnetic field $B$, for $\varepsilon=-2.5$ meV,
$A=4$ $\mu$eV, and $nhf=\Delta E$ for each $B$. In the whole $B$
range $q_{\mathrm{b}}/q_{\mathrm{d}}>10$ for $n=1$, and
$q_{\mathrm{b}}/q_{\mathrm{d}}>100$ for $n=2$ indicating an
appreciable speedup when the AC field tunes the tunnel coupling.
Near the anticrossing point $B\approx0.95$ T defined by $E_{1}$,
$E_{2}$ we have $h^{\mathrm{b}}_{11} - h^{\mathrm{b}}_{22}
\rightarrow 0$, hence $q_{\mathrm{b}} \rightarrow A
h^{\mathrm{b}}_{12}/2$ for $n=1$, but $q_{\mathrm{b}} \rightarrow
0$ for $n>1$; and similarly for $q_{\mathrm{d}}$.
Equation~(\ref{coupling}) shows that singlet-triplet transitions
vanish if $h^{\mathrm{b}}_{12}=0$, but this can occur far from the
negative detuning region of interest, and provided
$x_{\mathrm{so}}\ne 0$, due to the (anti-) bonding character of
the involved states.~\cite{suppl}

In the lower frame of Fig.~\ref{qdqb}
$q_{\mathrm{b}}/q_{\mathrm{d}}$ is plotted as a function of the
detuning for $A=4$ $\mu$eV. For an order of magnitude estimation
of $q_{\mathrm{b}}/q_{\mathrm{d}}$ the magnetic field is taken to
be 0.5 T higher than that of the anticrossing point. For large
negative detuning, $|S_{11}\rangle$ dominates over
$|S_{02}\rangle$ thus $q_{\mathrm{b}}/q_{\mathrm{d}}\gg 1$, which
indicates a significant speedup in the singlet-triplet transition
rate when the AC field tunes the tunnel coupling. For
$\varepsilon\rightarrow0$ $q_{\mathrm{b}}/q_{\mathrm{d}}$
decreases because the contributions of $|S_{11}\rangle$ and
$|S_{02}\rangle$ become gradually equal. Eventually, near zero
detuning the situation is more complicated and whether
$q_{\mathrm{b}}$ or $q_{\mathrm{d}}$ is greater depends on the
photon index $n$ and the magnetic field.~\cite{suppl}

To address possible implications in the experiment we consider the
scenario where the same AC voltage is applied to a
`tunnel-coupling gate' and separately to a `detuning gate', but
results in different AC field amplitudes, $A_{\mathrm{b}}$ and
$A_{\mathrm{d}}$ respectively. We define for the $n$-photon
resonance the effective ratio
$(q_{\mathrm{b}}/q_{\mathrm{d}})(A_{\mathrm{b}}/A_{\mathrm{d}})^n$,
where $q_{\mathrm{b}}/q_{\mathrm{d}}$ is given above and plotted
in Fig.~2 (lower frame). An effective ratio greater than unity
indicates that faster transitions are achieved by tuning with the
AC field the tunnel coupling. When
$A_{\mathrm{b}}/A_{\mathrm{d}}>1$ a larger speedup is achieved
compared to $A_{\mathrm{b}}=A_{\mathrm{d}}$, and the range of the
detuning in which the speedup can be probed is extended to smaller
absolute values. The opposite trends occur in the regime
$A_{\mathrm{b}}/A_{\mathrm{d}}<1$. However, even when
$A_{\mathrm{b}}/A_{\mathrm{d}}=0.1$ then for example at
$\varepsilon\approx-2.5$ meV, the speedup for the $n=1-3$ photon
resonances is between 2 and 5, and for the $n=4-6$ photon
resonances is between 10 and 30. The speedup is even better for
$n>6$ and/or larger negative detuning. In this work,
$\varepsilon/t_{\mathrm{c}}\approx 7$ for the maximum
$\varepsilon$ considered, but ratios on the order of 100 have been
probed.~\cite{hanson07, zwanenburg13}

So far we focused on the transitions between the lowest
eigenstates $|\psi_{1}\rangle$, $|\psi_{2}\rangle$ and found that
for large negative $\varepsilon$ a significant speedup is achieved
in the transition rate when the AC field tunes the inter-dot
tunnel coupling. The dependence of the DD eigenstates on the
detuning, and specifically the change of character from
$|S_{11}\rangle$ to $|S_{02}\rangle$ suggests the generalisation
of this result to transitions between the higher eigenstates
$|\psi_{4}\rangle$, $|\psi_{5}\rangle$, which are also coupled by
the SOI, provided the sign of $\varepsilon$ is reversed, namely
when $\varepsilon$ is large positive [Ref.~\onlinecite{suppl}].
The large detuning regime for both negative and positive values
has been investigated in a spin-blockaded DD with an AC driven
detuning.~\cite{stehlik14}

\begin{figure}
\includegraphics[width=6.5cm, angle=270]{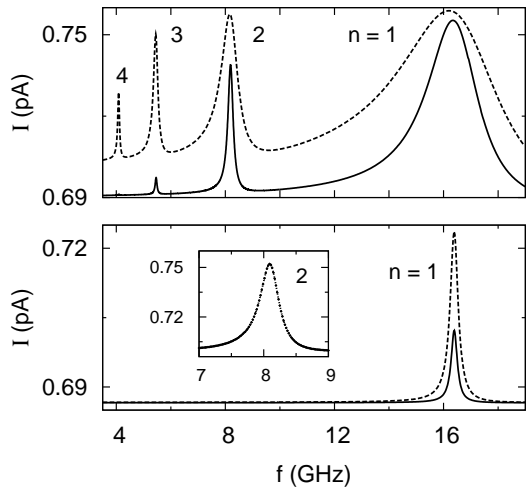}\\
\caption{Current as a function of AC field frequency. In the upper
(lower) frame the AC field tunes the inter-dot tunnel coupling
(energy detuning). The AC amplitude is $A=20$ $\mu$eV (40 $\mu$eV)
for solid (dashed) lines. $n=1$, 2, ... is the $n$-photon
resonance. The inset shows $n=2$ for $A=400$
$\mu$eV.}\label{curvsf}
\end{figure}

\section{AC-induced current}

To calculate the electrical current flowing through the DD in the
presence of the AC field we employ a Floquet-Markov master
equation approach.~\cite{flqmas1, flqmas2} The electrons in the
leads are described by the Hamiltonian $H_{\mathrm{e}}=\sum_{\ell,
k,\sigma}\epsilon_{\ell k}d^{\dagger}_{\ell k\sigma} d_{\ell
k\sigma}$ where the operator $d^{\dagger}_{\ell k \sigma}$
($d_{\ell k\sigma}$) creates (annihilates) an electron in lead
$\ell=\{$L, R$\}$, with momentum $k$, spin $\sigma$, and energy
$\epsilon_{\ell k}$. The tunnelling Hamiltonian accounts for the
interaction between the DD and the two leads
\begin{equation}\notag
H_{\mathrm{T}} = t_{\mathrm{T}}\sum_{
k,\sigma}(c_{1\sigma}^{\dagger}d_{\mathrm{L}
k\sigma}+c_{2\sigma}^{\dagger}d_{\mathrm{R} k\sigma})
+\text{H.c.},
\end{equation}
and $c^{\dagger}_{i\sigma}$ is the electron creation operator on
dot $i$ with spin $\sigma$. The tunnelling rates describing
tunnelling events in and out of the DD with a change in the
electron number by $\pm 1$, are calculated to second order in the
dot-lead tunnel coupling $t_{\mathrm{T}}$. The matrix elements
which are involved in the tunnelling rates are spin dependent and
account for spin blockade effects. We are interested in finding
the density matrix $\rho(t)$ of the DD, and in order to facilitate
the calculations we express $\rho(t)$ in the basis defined by the
Floquet modes $|u(t)\rangle$. In the steady-state we assume that
$\rho_{\mathrm{st}}(t)=\rho_{\mathrm{st}}(t+T)$, and
$\rho_{\mathrm{st}}(t)$ is expanded in a Fourier series allowing
for the equation of motion to be written in a matrix form and to
be solved numerically. Finally, using $\rho_{\mathrm{st}}(t)$ and
the tunnelling rates we calculate the time average current over a
AC period. The dot-lead coupling constant, proportional to
$t^{2}_{\mathrm{T}}$, is $\Gamma=1.2$ GHz ($\approx$ 5 $\mu$eV).
An important aspect is that unlike the two-level model that
examines the singlet-triplet transitions, the quantum transport
model considers all DD eigenstates to account correctly for the
various populations.~\cite{double, giavaras11} These are important
not only for the AC-induced peaks but also for the background
current~\cite{ono17, giavaras11} $I_{\mathrm{b}}$ for $A=0$.

Figure~\ref{curvsf} shows the current as a function of the AC
field frequency $f$, for different AC field amplitudes $A$, and
$B=1.5$ T, $\varepsilon=-2.5$ meV. Both the background and the
AC-induced currents are sensitive to these two parameters. In the
upper frame of Fig.~\ref{curvsf} the AC field tunes the inter-dot
tunnel barrier. Provided $A\ne0$ a current peak is formed when the
resonant condition $nhf\approx \Delta E = E_{2}-E_{1}$ is
satisfied, with $n=1$, 2, ... and $E_{i}$ being the two lowest
eigenenergies. Thus, the formation of the current peaks is a
result of singlet-triplet transitions caused by the AC field. The
single-photon peak ($n=1$) is the strongest, whereas multi-photon
peaks ($n>1$) are successively weaker. By increasing the amplitude
$A$ the current peaks become stronger because the transition rates
increase within the parameter range of this study, and for weak
driving the rates are proportional to $q^{2}_{\mathrm{b}}$.
However, the system is not driven through the anticrossing point
(upper frames Fig.~1) to undergo Landau-Zener dynamics. In the
chosen frequency range up to four peaks can be seen, but peaks for
$n>4$ can equally well be examined.

\begin{figure}
\includegraphics[width=4.cm,angle=270]{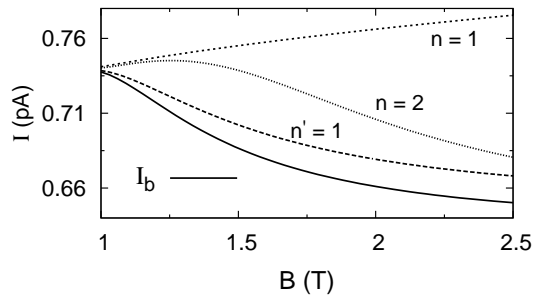}\\
\caption{Current on resonance at different magnetic fields for AC
amplitude $A=20$ $\mu$eV. $I_{\mathrm{b}}$ is the background
current for $A=0$. $n=1$, 2 is the $n$-photon resonance when the
AC field tunes the tunnel coupling, and $n'=1$ is the single
photon resonance when the AC field tunes the energy detuning. The
$n'=2$ resonance is vanishingly small and not
shown.}\label{curvsB}
\end{figure}

In the lower frame of Fig.~\ref{curvsf} the AC field tunes the
energy detuning, and similarly to the upper frame a current peak
is expected when $nhf\approx\Delta E$. However, the situation is
now different, and only the $n=1$ peak can be seen, whereas the
$n>1$ peaks are suppressed. In addition, the $n=1$ peak is much
weaker compared to the $n=1$ peak in the upper frame due to the
slower transition rate, which scales approximately as
$q^{2}_{\mathrm{d}}$. The suppression of the $n>1$ peaks results
from the small value of $A$; $n>1$ peaks are formed provided $A$
is large. For example, for $A=400$ $\mu$eV the $n=2$ peak in the
lower frame [inset to Fig.~\ref{curvsf}] becomes comparable to
that in the upper frame which corresponds to $A=40$ $\mu$eV. For
$n>2$ peaks to be formed the amplitude $A$ has to be even larger.
Charge noise arising from voltage fluctuations on the gate
electrodes can affect the AC peaks. However, the peaks can still
be probed even when $A$ is as large as 1.3 meV
[Ref.~\onlinecite{stehlik14}], and when the gate electrode design
is rather limited.~\cite{ono17} The impact of noise on the peaks
depends on the DD fabrication details.

Figure~\ref{curvsB} shows the resonant current (maximum value) at
different magnetic fields $B$, for $A=20$ $\mu$eV and
$\varepsilon=-2.5$ meV. The AC frequency corresponds to the
(approximate) resonant frequency at each $B$, i.e., $nhf = \Delta
E$. The chosen range of magnetic field $B>1$ T somewhat simplifies
the presentation, since for $B<1$ T various allowed transitions
between DD eigenstates lead to overlapping peaks. Near the
anticrossing point defined by $E_{1}$, $E_{2}$ ($B\approx 0.95$ T)
the resonant current in all cases is suppressed, and is
approximately equal to the background current $I_{\mathrm{b}}$.
The reason is that when $A=0$ the populations of the two
eigenstates forming the anticrossing are almost equal, therefore
applying the AC field has negligible effect.~\cite{ono17} Away
from the anticrossing the populations are different and the peaks
can increase, though the effect of the SOI decreases with $B$, so
a non monotonous behaviour can be observed. Within a simplified
approach, and assuming that the weak driving regime is a good
approximation the behaviour depends on the ratio
$q^{2}_{\mathrm{b}}/\Gamma$. If it is large the peak starts to
decrease at a high magnetic field. When the AC field tunes the
tunnel barrier the $n=1$ peak increases with $B$, while the $n=2$
peak first increases and then starts to decrease; the relative
peak height (measured with respect to $I_{\mathrm{b}}$) is maximum
at about 1.6 T. This large difference in the two behaviours is due
to the fact that $q_{\mathrm{b}}(n=1)\gg q_{\mathrm{b}}(n=2)$. The
increase of the $n=1$ peak occurs even at relatively high $B$,
because the populations are very different, and according to
Fig.~\ref{qdqb} $q_{\mathrm{b}}$ decreases slowly with $B$ in the
high $B$ regime. In contrast, as shown in Fig.~\ref{curvsB} when
the AC field tunes the detuning only the $n=1$ peak can be clearly
observed, and this is now much weaker because $q_{\mathrm{b}}\gg
q_{\mathrm{d}}$.

\section{Conclusion}

We demonstrated that when the AC field tunes the inter-dot tunnel
coupling the singlet-triplet transitions can be over an order of
magnitude faster compared to the case where the AC field tunes the
detuning. As a result, the AC field induced current peaks are
well-formed at much smaller AC amplitude. Multi-photon resonances
are enhanced by orders of magnitude allowing for spectroscopy at
smaller frequencies. Our findings are useful for quantum dot spin
qubits where fast operations are needed with small AC amplitudes
and frequencies.

\setcounter{secnumdepth}{0} 

\section{Acknowledgement}

Part of this work was supported by CREST JST (JPMJCR15N2).


\setcounter{secnumdepth}{1}

\newpage

\setcounter{secnumdepth}{0} 

\section{Supplemental material for: Spectroscopy of double
quantum dot two-spin states by tuning the inter-dot barrier}

\setcounter{secnumdepth}{1}

\setcounter{equation}{0} \setcounter{figure}{0}
\setcounter{table}{0} \setcounter{page}{1} \makeatletter

\setcounter{section}{0}

\renewcommand{\theequation}{S\arabic{equation}}
\renewcommand{\thefigure}{S\arabic{figure}}
\renewcommand{\bibnumfmt}[1]{[S#1]}
\renewcommand{\citenumfont}[1]{S#1}





\section{Two-electron energy spectrum}

In the singlet-triplet basis $|S_{11}\rangle$, $|S_{02}\rangle$,
$|T_{0}\rangle$, $|T_{+}\rangle$, $|T_{-}\rangle$ the double dot
Hamiltonian is
\begin{equation}
\begin{split}
&H_{\mathrm{DD}}=\Delta[|T_-\rangle\langle T_-|-|T_+\rangle\langle
T_+|]-\delta|S_{02}\rangle\langle
S_{02}|\\
&-\sqrt{2}T_{\mathrm{c}}|S_{11}\rangle\langle S_{02}| -
T_{\mathrm{so}}[|T_+\rangle\langle S_{02}|+ |T_-\rangle\langle
S_{02}|]+ \text{H.c.}
\end{split}
\end{equation}
Here, the singlet state $|S_{20}\rangle$ is ignored but in all the
numerical computations this is taken into account. The
eigenenergies and eigenstates of the two electrons are computed by
solving the eigenvalue problem $H_{\mathrm{DD}} |\psi_i\rangle =
E_{i}|\psi_i\rangle$. Only the time independent part of
$H_{\mathrm{DD}}$ is considered and the AC field amplitude is
$A=0$. For simplicity, we refer to $|\psi_i\rangle$ as singlet and
triplet eigenstates even when the spin-orbit interaction is
nonzero, and we label $|\psi_i\rangle$ in order of increasing
eigenenergy $E_{i}$.

\begin{figure}[b]
\begin{center}
\includegraphics[width=7 cm, angle=270]{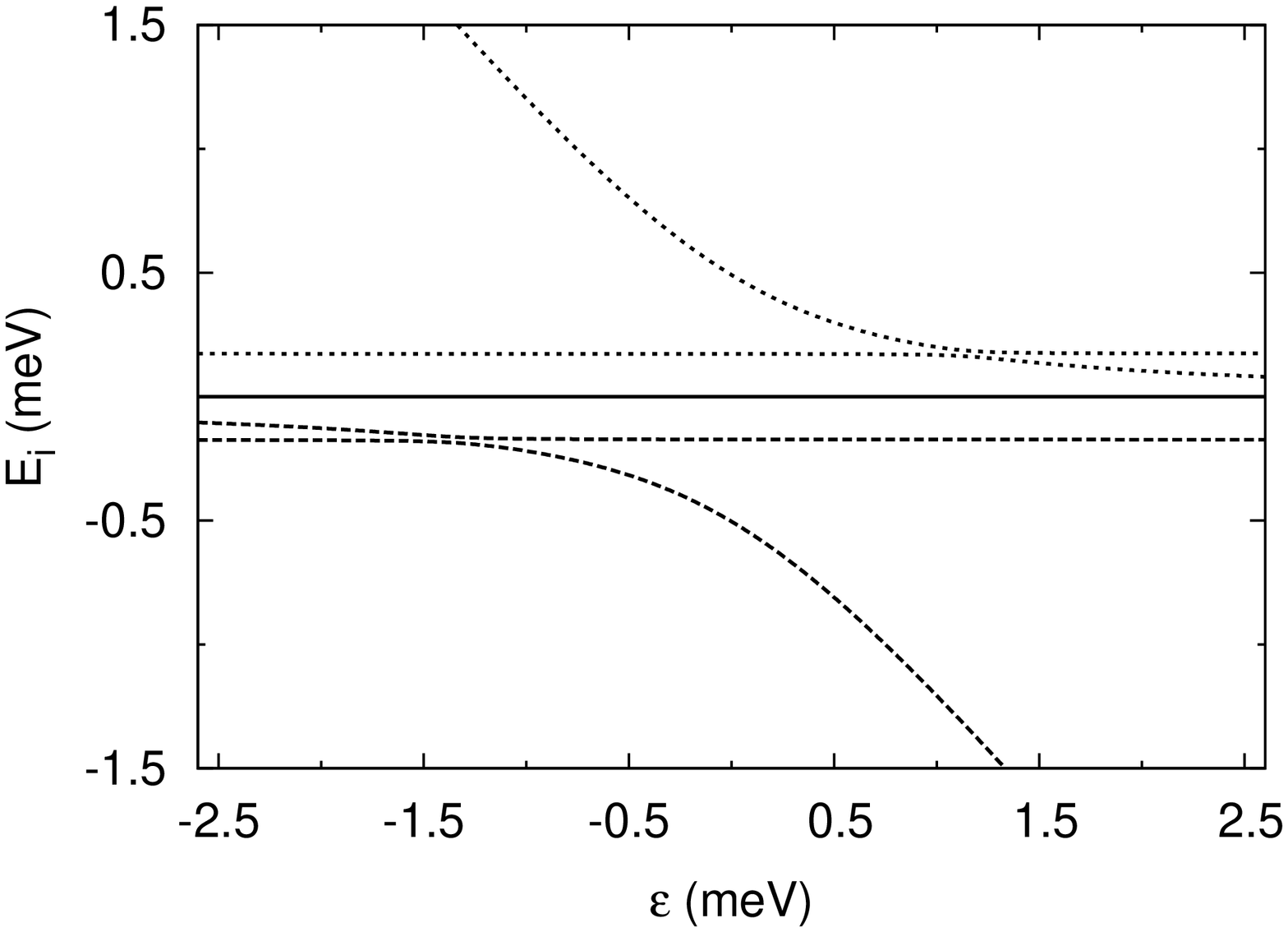}
\caption{Eigenenergies as a function of energy detuning. The
eigenenergies $E_{1}$, $E_{2}$ are plotted by dashed lines, and
the eigenenergies $E_{4}$, $E_{5}$ are plotted by dotted
lines.}\label{S1}
\end{center}
\end{figure}

Figure~\ref{S1} shows the eigenenergies $E_{i}$ as a function of
the detuning ($\delta=\varepsilon$ for $A=0$) for the same
parameters as those used in the main article $B=1.5$ T,
$t_{\mathrm{c}}=0.35$ meV, $t_{\mathrm{so}}=0.035$ meV. The two
anticrossing points for $\varepsilon \ne 0$ are formed by the
spin-orbit interaction, and the anticrossing point for
$\varepsilon=0$ is formed by the inter-dot tunnel coupling. The
eigenenergies $E_{1}$, $E_{2}$ are plotted by dashed lines (see
also Fig.~1 in the main article) and in the main article we
examine the transitions between $|\psi_{1}\rangle$ and
$|\psi_{2}\rangle$. In Sec.~\ref{higher} of the Supplemental
Material we generalise our findings to the transitions between
$|\psi_{4}\rangle$ and $|\psi_{5}\rangle$ whose eigenenergies
$E_{4}$, $E_{5}$ are plotted in Fig.~\ref{S1} by dotted lines.


\section{Approximate two-level Hamiltonian}

In the main article the transitions between the two lowest
singlet-triplet eigenstates $|\psi_{1}\rangle$, $|\psi_{2}\rangle$
are studied within the Floquet formalism. In addition an
approximate two-level Hamiltonian is shown to predict the correct
features. The main advantage of this Hamiltonian is that it is
time independent and its derivation follows the same methodology
as that presented in Ref.~\onlinecite{ono2017}. For completeness
we briefly outline the basic steps of the derivation here. Because
we are interested in the transitions between the two lowest
eigenstates $|\psi_1\rangle$, $|\psi_2\rangle$ whose levels
$E_{1}$, $E_{2}$ anticross, it is convenient to write the total DD
Hamiltonian in the energy basis $|\psi_i\rangle$ using the
notation
\begin{equation}
H_{\mathrm{DD}} = h_{\mathrm{E}} + A\sin(2 \pi f t)h^{j}.
\end{equation}
The matrix $h_{\mathrm{E}}$ is diagonal with elements $E_{i}$, and
the matrix $h^{j}$ contains the time independent `coupling'
constants due to the AC field. For the matrix $h^{j}$ we have to
distinguish the two different cases for the AC field: when the AC
field tunes the tunnel coupling leading to a time dependent
$T_{\mathrm{c}}$, $T_{\mathrm{so}}$ then $h^{j}=h^{\mathrm{b}}$,
and when the AC field tunes the energy detuning leading to a time
dependent $\delta$ then $h^{j}=h^{\mathrm{d}}$. When the AC field
frequency satisfies $nh f \approx E_{2}-E_{1}$, $n=1, 2,$ ... and
the AC amplitude is small prohibiting transitions to levels
$E_{i}$, $i>2$, we can assume that the dynamics is restricted
solely within $E_{1}$, $E_{2}$. Under these conditions and for the
case of $h^{\mathrm{b}}$ we can focus on the following $2\times2$
Hamiltonian
\begin{equation}
h_{\mathrm{DD}}=\left(\begin{array}{cc}
   E_{1}+A\sin(2 \pi f
t)h^{\mathrm{b}}_{11} & A\sin(2 \pi f
t)h^{\mathrm{b}}_{12} \\
   A\sin(2 \pi f
t)h^{\mathrm{b}}_{21} & E_{2}+A\sin(2 \pi f
t)h^{\mathrm{b}}_{22} \\
 \end{array}\right),
\end{equation}
where $h^{\mathrm{b}}_{nm}$ denotes matrix elements of
$h^{\mathrm{b}}$. These elements can be easily calculated and are
given in Eq.~(2) in the main article. The next step is to
transform $h_{\mathrm{DD}}$ by applying the operator
\begin{equation}
U=\left(%
\begin{array}{cc}
  e^{i\phi_1(t)} & 0  \\
  0 & e^{i\phi_2(t)}  \\
\end{array}%
\right),
\end{equation}
and choosing $\phi_i(t)$ to remove the time dependence from the
diagonal elements of $h_{\mathrm{DD}}$ as well as to introduce the
`photon' shift $nhf$ with $n=1, 2$, ... The transformed
Hamiltonian is
\begin{equation}
W = \left(\begin{array}{cc}
   E_{1}+nhf/2 & W_{12} \\
   W^{*}_{12} & E_{2}-nhf/2 \\
\end{array}\right),
\end{equation}
with
\begin{equation}\label{Wfull}
W_{12} = \frac{A}{2}h^{\mathrm{b}}_{12} \left(e^{i(-n+1) 2\pi f t}
- e^{i(-n-1)2\pi f t} \right) \sum^{\infty}_{m=-\infty} i^{3m-1}
e^{i m 2\pi f t} J_{m}\left(
\frac{A(h^{\mathrm{b}}_{11}-h^{\mathrm{b}}_{22})}{hf}\right),
\end{equation}
and $J_{m}$ is the 1st kind Bessel function of order $m$.

Because $|h^{\mathrm{b}}_{11}-h^{\mathrm{b}}_{22}|<1$, and
$|h^{\mathrm{b}}_{12}|<1$ the dynamics in the regime $A\ll hf$ can
be described approximately by retaining only the time independent
terms
\begin{equation}
W_{12} \approx \frac{n h f
h^{\mathrm{b}}_{12}}{(h^{\mathrm{b}}_{11} - h^{\mathrm{b}}_{22})}
J_{n}\left(\frac{A(h^{\mathrm{b}}_{11} - h^{\mathrm{b}}_{22})}{hf}
\right) = q_{\mathrm{b}}.
\end{equation}
Here, we ignore a factor of $i^{3n}$ and the derived
$q_{\mathrm{b}}$ is the coupling constant given in Eq.~(1) in the
main article. The approximation is sensitive to the size of the
quantities $h^{\mathrm{b}}_{11}-h^{\mathrm{b}}_{22}$, and
$h^{\mathrm{b}}_{12}$. These depend on the double dot parameters
(detuning, inter-dot tunnel coupling) as well as the magnetic
field, and if these quantities are made small then
$q_{\mathrm{b}}$ can still be employed for $A\sim hf$. A
particular case occurs at the anticrossing point where
$h^{\mathrm{b}}_{11}-h^{\mathrm{b}}_{22} \rightarrow 0$ and
$W_{12}$ in Eq.~(\ref{Wfull}) has no time independent terms for
$n>1$. In the present work this case is not interesting, because
the AC induced current near the anticrossing point is suppressed
(Fig.~4 in the main article) due to the vanishingly small
population difference of the two eigenstates forming the
anticrossing.

Finally, defining $\Delta E=E_{2}-E_{1}$ the form of the
Hamiltonian $W$ is
\begin{equation}
W = \left(\begin{array}{cc}
   -(\Delta E-nhf)/2 & q_{\mathrm{b}} \\
   q_{\mathrm{b}} & (\Delta E-nhf)/2 \\
\end{array}\right) = - (\Delta E-nhf)/2\sigma_z + q_{\mathrm{b}}
\sigma_x.
\end{equation}
In the main article, the approximate singlet-triplet transition
probability $P_{\mathrm{ap}}$ is derived from this Hamiltonian.

\section{Inspection of coupling constants $q_{\mathrm{b}}$ and $q_{\mathrm{d}}$}

\begin{figure}
\begin{center}
\includegraphics[width=6 cm, angle=270]{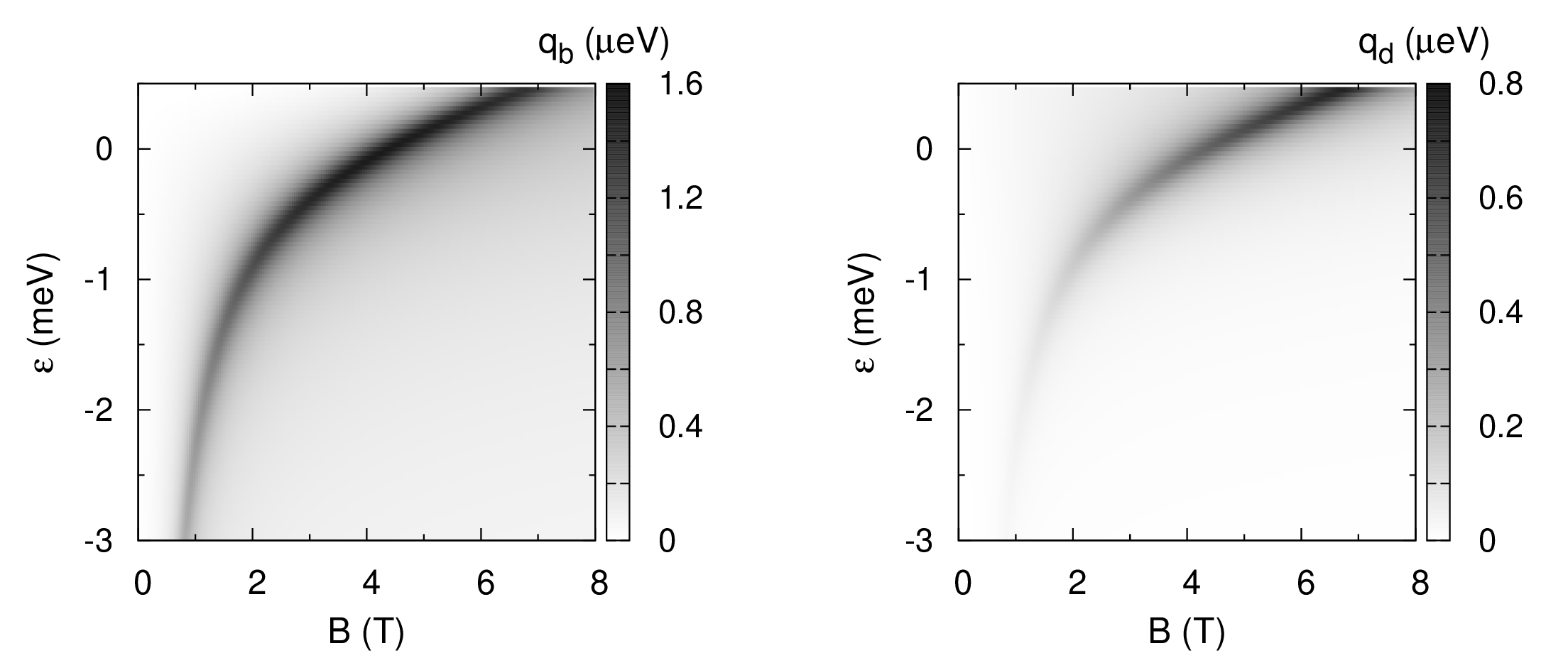}
\caption{Coupling constants $q_{\mathrm{b}}$ and $q_{\mathrm{d}}$
as a function of energy detuning and magnetic field.}\label{S2}
\end{center}
\end{figure}

Figure~\ref{S2} shows the coupling constants $q_{\mathrm{b}}$ and
$q_{\mathrm{d}}$ (we consider everywhere the absolute values) as a
function of the energy detuning and magnetic field for the AC
field amplitude $A=4$ $\mu$eV and $n=1$. In both cases the maximum
occurs near the anticrossing point which shifts at higher field as
the detuning increases. The large negative detuning regime, where
the spins are effectively in the Heisenberg limit, is of
particular interest in this work when we focus on the transitions
between $|\psi_{1}\rangle$ and $|\psi_{2}\rangle$. In this regime
$q_{\mathrm{b}} \gg q_{\mathrm{d}}$ suggesting that
singlet-triplet transitions are much faster when the AC field
tunes the inter-dot tunnel coupling instead of the energy
detuning. Specifically, for $\varepsilon \lesssim -1.4$ meV the
transitions can be over an order of magnitude faster. Near zero
detuning $q_{\mathrm{b}}$ and $q_{\mathrm{d}}$ are on the same
order of magnitude.

\begin{figure}
\begin{center}
\includegraphics[width=5.5 cm, angle=270]{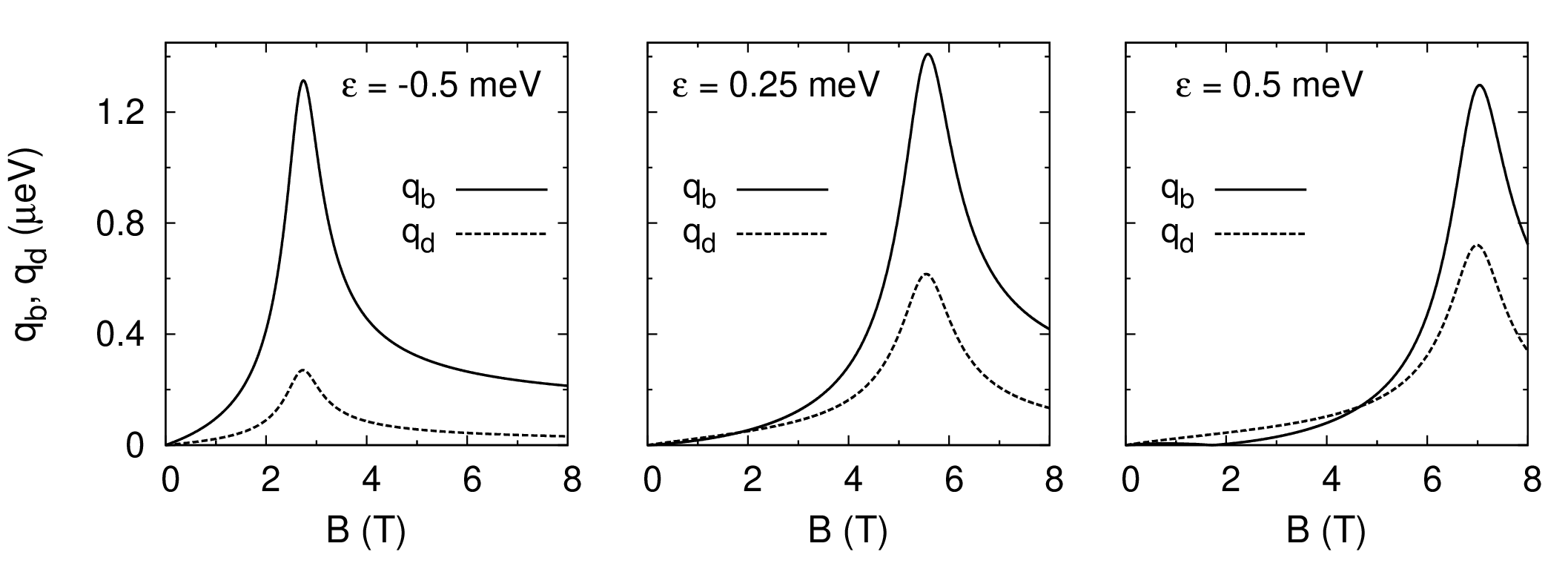}
\caption{Coupling constants $q_{\mathrm{b}}$ and $q_{\mathrm{d}}$
as a function of magnetic field at fixed energy
detuning.}\label{S3}
\end{center}
\end{figure}

In Fig.~2 of the main article the coupling constants
$q_{\mathrm{b}}$ and $q_{\mathrm{d}}$ are plotted as a function of
the magnetic field for large negative detuning $\varepsilon =
-2.5$ meV. Figure~\ref{S3} shows $q_{\mathrm{b}}$ and
$q_{\mathrm{d}}$ at different detunings for the AC field amplitude
$A=4$ $\mu$eV and $n=1$. For positive detuning $q_{\mathrm{d}}$
can be equal or greater than $q_{\mathrm{b}}$. Moreover, for
$\varepsilon = 0.5$ meV $q_{\mathrm{b}}=0$ at finite magnetic
field $B\approx1.8$ T. The reason for this particular behaviour is
that $h^{\mathrm{b}}_{12}=0$ because $x_{\mathrm{so}}\ne0$ and the
coefficients in Eq.~(2) in the main article have different signs
due to the (anti-) bonding character of the states. However, in
the large negative detuning regime and assuming that
$x_{\mathrm{so}}$ is on the order of
$t_{\mathrm{so}}/t_{\mathrm{c}}$ then $h^{\mathrm{b}}_{12}\ne0$,
because either positive or negative terms dominate. In contrast,
$h^{\mathrm{d}}_{12}\ne0$ for any value of the detuning.

\section{Transitions between higher eigenstates}\label{higher}

\begin{figure}
\begin{center}
\includegraphics[width=5 cm, angle=270]{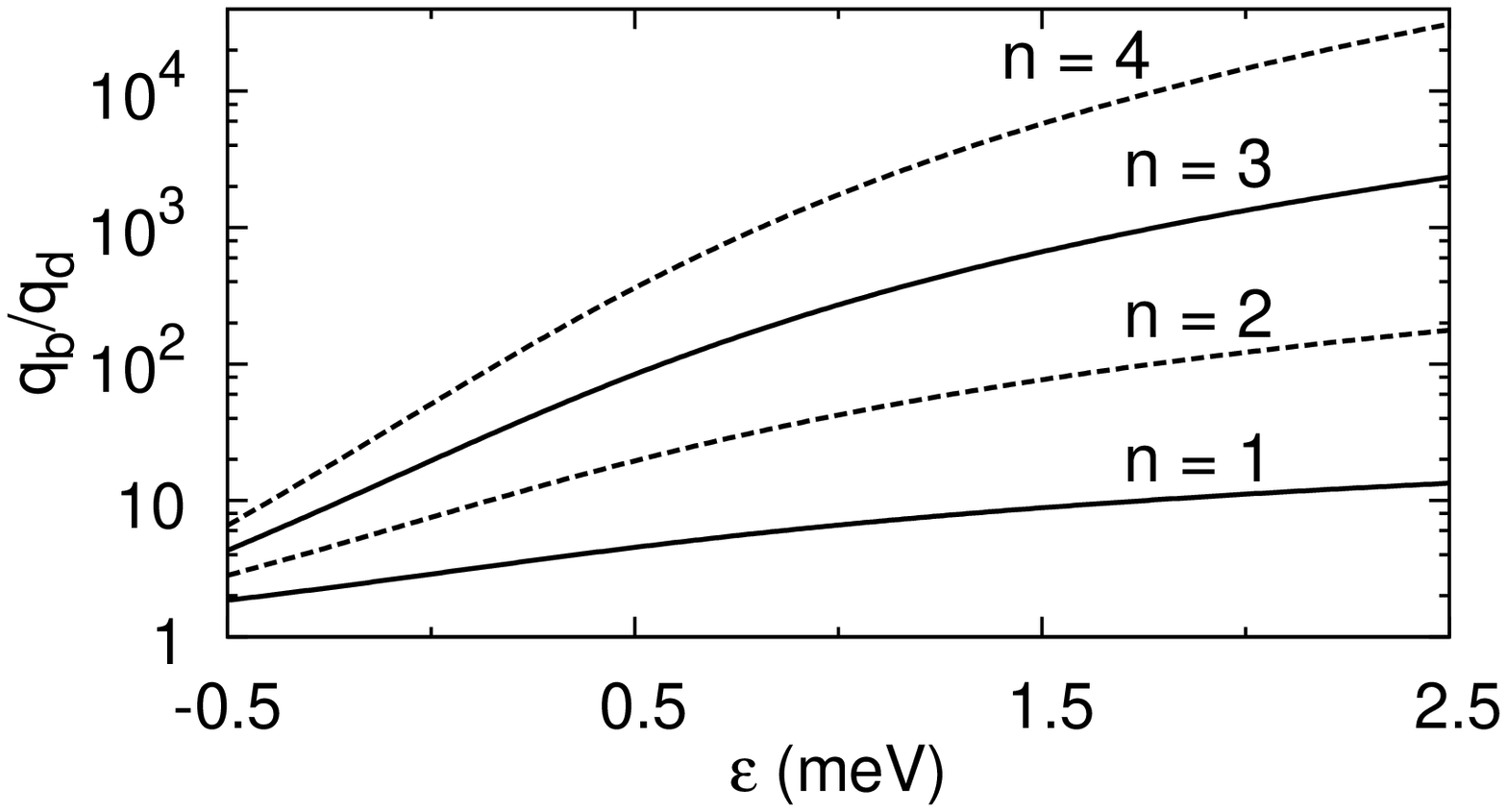}
\caption{The figure shows the ratio
$q_{\mathrm{b}}/q_{\mathrm{d}}$ on logarithmic scale as a function
of energy detuning for the $n=1-4$ $n$-photon resonance. Here, the
two eigenstates correspond to $|\psi_{4}\rangle$,
$|\psi_{5}\rangle$.}\label{S4}
\end{center}
\end{figure}

In the main article the transitions between the two lowest
singlet-triplet eigenstates $|\psi_{1}\rangle$, $|\psi_{2}\rangle$
are examined. However, the derivation of the two-level Hamiltonian
$W$ suggests that $W$ can describe equally well the transitions
between the two eigenstates $|\psi_{4}\rangle$, $|\psi_{5}\rangle$
by simply considering the appropriate matrix elements which couple
$|\psi_{4}\rangle$ to $|\psi_{5}\rangle$ and calculating the
coupling constants $q_{\mathrm{b}}$, $q_{\mathrm{d}}$. The
eigenenergies $E_{4}$, $E_{5}$ are shown in Fig.~\ref{S1}.

We focus on the singlet-triplet eigenstates $|\psi_{4}\rangle$,
$|\psi_{5}\rangle$ and plot in Fig.~\ref{S4} the ratio
$q_{\mathrm{b}}/q_{\mathrm{d}}$ as a function of the energy
detuning for the AC amplitude $A=4$ $\mu$eV, and $n=1-4$. As in
the main article, the magnetic field is taken to be 0.5 T higher
than that of the anticrossing point. Following the arguments given
in the main article we expect $q_{\mathrm{b}}/q_{\mathrm{d}}\gg 1$
when the $|S_{11}\rangle$ component dominates over
$|S_{02}\rangle$. This condition is now satisfied for large
positive detuning. According to Fig.~\ref{S4}, when the AC field
tunes the inter-dot tunnel coupling it results in a significant
speedup in the transition rate between the singlet-triplet
eigenstates $|\psi_{4}\rangle$, $|\psi_{5}\rangle$. This speedup
is of the same order of magnitude as that achieved for the
eigenstates $|\psi_{1}\rangle$, $|\psi_{2}\rangle$ examined in the
main article (see Fig.~2 lower panel). The basic difference is in
the range of the energy detuning. In particular, when the AC field
tunes the tunnel coupling the transitions between the
singlet-triplet eigenstates $|\psi_{4}\rangle$, $|\psi_{5}\rangle$
are over an order of magnitude faster for the detuning
$\varepsilon \gtrsim +1.6$ meV.




\begin{thebibliography}{10}
\bibitem{com1}
Two-spin eigenstates are singlet-triplet superpositions, but we
refer to singlet and triplet states for simplicity.
\bibitem{extranote}
The choice of g-factor is not important, but it determines the
required magnetic field to probe the singlet-triplet anticrossing
\bibitem{absol}
The absolute values of $q_{\mathrm{b}}$, $q_{\mathrm{d}}$ are
considered everywhere.
\end{thebibliography}
\end{document}